\documentclass[usenatbib]{mn2e}
\usepackage{natbib}
\usepackage{amssymb}
\usepackage{epsfig}
\usepackage{rotating} 

\newcommand{\Chandra}{\emph {Chandra}\ }
\newcommand{\XMM}{\emph{XMM-Newton}\ }
\newcommand{\Einstein}{\emph{Einstein}\ }
\newcommand{\keV}{\mbox{\ensuremath{\mathrm{~keV}}}}
\newcommand{\LCDM}{$\Lambda$CDM~}
\newcommand{\erg}{\mbox{\ensuremath{\mathrm{~erg}}}}
\newcommand{\s}{\mbox{\ensuremath{\mathrm{~s}}}}
\newcommand{\ps}{\ensuremath{\mathrm{\s^{-1}}}}
\newcommand{\ergps}{\ensuremath{\mathrm{\erg \ps}}}
\newcommand{\pcmsq}{\mbox{\ensuremath{\mathrm{~cm^{-2}}}}}

\def\r200{\ensuremath{r_{\mathrm{200}}}}
\newcommand{\egc}{{\it e.g.}}  
\newcommand{\eg}{{\it e.g.\ }}
\newcommand{\chisq}{\ensuremath{\chi^2}}
\newcommand{\ks}{\mbox{\ensuremath{\mathrm{~ks}}}}
\newcommand{\Zsol}{\ensuremath{\mathrm{Z_{\odot}}}}
\newcommand{\arcm}{\ensuremath{\mathrm{^\prime}}}
\newcommand{\arcs}{\arcm\hskip -0.1em\arcm}
\newcommand{\ROSAT}{\emph{ROSAT}\ }
\newcommand{\MEKAL}{\textsc{MeKaL}\ }
\newcommand{\km}{\mbox{\ensuremath{\mathrm{~km}}}}
\newcommand{\Mpc}{\mbox{\ensuremath{\mathrm{~Mpc}}}}
\newcommand{\pMpc}{\ensuremath{\mathrm{\Mpc^{-1}}}}
\newcommand{\kmpspMpc}{\ensuremath{\mathrm{\km \ps \pMpc\,}}}
\newcommand{\kpc}{\mbox{\ensuremath{\mathrm{~kpc}}}}
\newcommand{\ie}{{\it i.e.\ }}
\newcommand{\etal}{{\it et al.\thinspace}}
\newcommand{\ctpp}{\ensuremath{\mathrm{ ~counts ~pixel^{-1}}}}
\newcommand{\ctps}{\ensuremath{\mathrm{ ~counts \ps}}}
\newcommand{\flux}{\ensuremath{\mathrm{\erg \ps \pcmsq}}}
\newcommand{\cm}{\mbox{\ensuremath{\mathrm{~cm}}}}
\newcommand{\cmsq}{\ensuremath{\mathrm{\cm^2}}}
\newcommand{\ent}{\ensuremath{\mathrm{\keV \cmsq}}}
\newcommand{\kmps}{\ensuremath{\mathrm{\km \ps}}}
\newcommand{\Gyr}{\mbox{\ensuremath{\mathrm{~Gyr}}}}

\begin{document}

\title[An \XMM observation of ClJ1226.9$+$3332]{An \XMM observation of the
  massive, relaxed galaxy cluster ClJ1226.9$+$3332 at $z=0.89$} 
\author[B. J. Maughan \etal]
  {B. J. Maughan,$^1$\thanks{E-mail: bjm@star.sr.bham.ac.uk}
    L. R. Jones,$^1$ H. Ebeling $^2$ and C. Scharf $^3$\\ 
  $^1$School of Physics and Astronomy, The University of Birmingham,
  Edgbaston, Birmingham B15 2TT, UK\\
  $^2$Institute for Astronomy, 2680 Woodlawn Drive, Honolulu, HI 96822,
  USA\\ 
  $^3$Columbia Astrophysics Laboratory, MC 5247, 550 West 120th St., New York, NY 10027, USA\\
}

\maketitle

\begin{abstract}
A detailed X-ray analysis of an \XMM observation of the high-redshift
(z=0.89) galaxy cluster ClJ1226.9$+$3332 is presented. After careful
consideration of background subtraction issues, the X-ray temperature
is found to be $11.5^{+1.1}_{-0.9}\keV$, the highest X-ray temperature
of any cluster at $z>0.6$. The temperature is consistent with
the observed velocity dispersion. In contrast to MS1054-0321, the only other very hot cluster currently
known at $z>0.8$, ClJ1226.9$+$3332 features a relaxed X-ray morphology,
and its high overall gas temperature is not caused by one or several
hot spots. The system thus constitutes a unique example of a
high redshift (z$>$0.8), high temperature (T$>$10 keV), relaxed
cluster, for which the usual hydrostatic equilibrium assumption, and the X-ray mass is most reliable.

A temperature
profile is constructed (for the first time at this redshift) and is
consistent with the cluster being isothermal out to $45\%$ of the
virial radius. Within the virial radius (corresponding to a measured
overdensity of a factor of 200), a total mass of
$1.4\pm0.5\times10^{15}M_\odot$ is derived, with a gas mass
fraction of $12\pm5\%$ (for a \LCDM cosmology and H$_0$=70
km s$^{-1}$ Mpc$^{-1}$). This total mass is similar to that of the
Coma cluster. The bolometric X-ray
luminosity is $5.3^{+0.2}_{-0.2}\times10^{45}\ergps$. Analysis of a short
{\it Chandra} 
observation confirms the lack of significant point-source
contamination, the temperature, and the luminosity, albeit with lower
precision. The probabilities of finding a cluster
of this mass within the volume of the discovery X-ray survey are $\sim8\times 10^{-5}$
for $\Omega_M=1$ and $0.64$ for $\Omega_M=0.3$, making $\Omega_M=1$ highly
unlikely.

The entropy profile suggests that entropy evolution is being observed.
The metal abundance (of $Z=0.33^{+0.14}_{-0.10}\Zsol$), gas mass
fraction, and gas distribution are consistent with those of local clusters; thus the bulk 
of the metals were in place by z=0.89.

\end{abstract}


\section{Introduction} \label{sect:intro}
Massive galaxy clusters form from the high-sigma tail of the initial
cosmological density distribution. As a result they are rare, but also very
powerful probes of cosmology. Given an assumed initial density
distribution, the properties of the massive cluster population can be
predicted under alternate cosmologies, and those predictions tested with
observations. The predictions of different cosmologies diverge with
redshift, making high-redshift, massive clusters the most useful objects to
distinguish between them.  

X-ray observations of galaxy clusters provide a useful means of measuring
their properties. The intra-cluster gas is extremely luminous in X-rays,
and measurements of the gas temperature and density distributions allow the
total mass of the system (the property most directly related to
cosmological predictions) to be inferred. This inference requires, however,
that the intra-cluster medium (ICM) be in hydrostatic equilibrium. If, as
believed, clusters form through a series of hierarchical mergers, then this
will only be the case some time after the last merger. Thus the most useful
objects to study for the purpose of constraining cosmological models in
this way, are high-redshift, massive, relaxed clusters of galaxies. These
are extremely rare. 

However the question of when a cluster can be considered to be relaxed is
something of a contentious issue. In a study of $368$ low$-z$ clusters
observed by \Einstein, \citet{jon99} found $\approx40\%$ to have
substructure in their X-ray images. This fraction is likely to be an
underestimate, as \Einstein was unable to resolve small scale
substructure. More recently, the high resolving power of \Chandra has
revealed substructure in clusters that were previously considered to be
relaxed, such as A1795 \citep{fab01,mar01}, and MS 1455.0+2232
\citep{maz02a}. On the other hand, X-ray derived masses of clusters that
appear relaxed in \Chandra observations have been found to agree well with
independent weak lensing mass measurements, at least in the inner regions
\citep[][and references therein]{all01b}. 

While \Chandra can accurately probe the gas properties in the central
regions of clusters, the strength of \XMM lies in its large collecting
area, which allows it to trace the gas density and temperature structure
out into the low surface-brightness emission at large radii, even at high
redshifts. This minimises the uncertainties involved in extrapolating these
properties out to the virial radius when deriving the total mass of the
system. The mass composition of massive galaxy clusters (\eg the baryonic
to total mass fraction) is believed to be representative of the universe as
a whole, due to their large size \citep[\egc][]{all02a}. Thus by directly
observing the ICM out to large radii, one obtains a more representative
measurement of these properties. 

In June 2001, \XMM made a $30\ks$ observation of galaxy cluster
ClJ1226.9$+$3332, one of the most distant, luminous clusters found in the
WARPS X-ray selected survey \citep{sch97,jon98a,ebe00,per02}. The cluster
was positioned $\approx4\arcm$ off-axis in order to investigate other
candidate clusters in the field, to be described in a future paper. The
discovery \ROSAT data indicated a high X-ray luminosity, but were
insufficient to accurately determine morphology or temperature. Optical
follow up found the cluster's redshift to be $0.89$ \citep{ebe01},
corresponding to a look back time of over half of the age of the universe,
and Sunyaev-Zel'dovich effect imaging confirmed that the cluster is both
hot and massive \citep{joy01}. We present here the results of a detailed
analysis of the \XMM data. There is also a fairly short ($10\ks$) archived
\Chandra observation of ClJ1226.9$+$3332, which has been examined by
\citet{cag01}. We have analysed these data in a way consistent with our \XMM
analysis in order to check consistency, and we draw comparisons at several
relevant points.  

Throughout this paper, a cosmology of $H_0=70\kmpspMpc$, and
$\Omega_M=0.3$ ($\Omega_\Lambda=0.7$) is adopted, unless stated otherwise,
and all errors are quoted at the $68\%$ level. At the cluster's redshift,
$1\arcs$ corresponds to $7.8\kpc$ in this cosmology. The virial radius
($\r200$) is defined as the radius within which the mean density is $200$
times the critical density at the redshift of observation. 

\section{Data Preparation} \label{sect:dataprep}
\begin{figure}
\begin{center}
\scalebox{0.4}{\includegraphics*{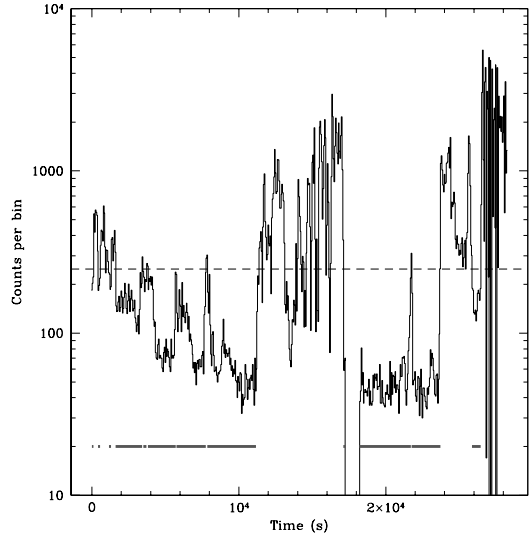}} \\
\caption{\label{fig:lightcurve}Lightcurve of the pn observation of
  ClJ1226.9$+$3332, in $50\s$ bins in the range $10-15\keV$. The bar below
  the lightcurve indicates the good time 
  intervals left after cleaning, and the dashed line indicates the
  $3\sigma$ cut-off level (see text).} 
\end{center}
\end{figure}

The data from the PN and two MOS detectors were processed with the
processing chains, epchain (PN) and emchain (MOS) as these have been found
to be significantly better at removing bad events and pixels than the
standard `procs' (epproc and emproc). Examination of the processed PN
events showed that a few bad pixels (two rows, and one pixel) were
undetected by the chain, and these were added to the bad pixel tables, and
the data was reprocessed. Lightcurves of the three detectors, produced in
the $10-15\keV$ band showed that the observation was contaminated by
several large background flares. The periods of very high background were
selected by eye, and removed from the lightcurve, before the remaining data
were cleaned by a recursive $3-\sigma$ clipping algorithm to leave a stable
mean rate. The lightcurve of the PN detector is shown in
Fig. \ref{fig:lightcurve}, with the accepted times indicated by the bar
underneath the lightcurve. 

Events were filtered on the basis of their pattern parameter, which
indicates the geometry of the detection of each event, \ie the number of
adjacent pixels that detect each photon. Events whose patterns are
considered well calibrated (PN - single and double, MOS - single, double,
and quadruple) were retained in the filtering. 

In the analysis of \XMM data, one must carefully account for the background
contamination. There are, broadly speaking, two ways of doing this; one may
sample the background locally, from the same observation as the source, or
one may use a `blank-sky' background dataset, composed of many
observations, with all bright sources removed \citep{lum02}.  

The background is composed of three types of events:
\begin{itemize}
\item Soft protons - this component is believed to be caused by solar
  flares, and the intensity and spectrum of this component varies
  significantly with time. It is the dominant component during flaring
  intervals, but in quiescent periods its contribution is the
  smallest. This component is vignetted, but may not have the same
  vignetting function as the X-rays. 
\item X-rays - this component dominates the background at low energies ($<1.5\keV$), and varies spatially across the sky (though not significantly across the field-of-view). This component is vignetted by the telescopes.
\item Cosmic-ray induced particles - this component dominates at high
  energies, and is induced by high energy cosmic rays passing unvignetted
  through the instrument. This component is referred to hereafter as the
  particle background. 
\end{itemize}

This observation of ClJ1226.9$+$3332 appears to be contaminated by a
particularly high background level, even after lightcurve cleaning. As
shown in Fig. \ref{fig:lightcurve}, there are two intervals of lower
background separated by a large flaring event. During the first interval
($2-11\ks$), for the PN camera, the average count rate ($10-15\keV$) was
$1.81\ctps$, while in the second($18-26\ks$), the mean rate was
$0.97\ctps$. For comparison, the PN count rate in the blank-sky datasets in
this energy band was $0.52\ctps$. Even considering the $10-20\%$ variations
in the background level found by \citet{lum02}, the background in these two
periods is higher than would normally be acceptable. The count rates
outside the field of view, which consist only of particle events, were also
compared. The count rate was a factor of $1.7$ higher in the
ClJ1226.9$+$3332 dataset than the blank-sky data. This shows that the high
background level in this dataset is due to high levels of both particles
and soft protons. This significantly increases the difficulty and
uncertainties involved with using a blank-sky background in the analysis.  

Due to the high background, a careful comparison of spectral analysis
methods was made (using both local and blank-sky backgrounds), on both of
the time periods separately, and combined (for simplicity, hereafter, the
first period ($2-11\ks$) will be referred to as the ``high background
period'', and the second ($18-26\ks$) will be referred to as the ``low
background period''). This analysis is described in some detail in the
following sections, but the general conclusion was that in the low
background period, all methods gave consistent results, and that if a local
background was used, then the results from the high background and low
background periods, and both periods combined were consistent. As discussed
in \textsection \ref{sec:specsum} our final results are taken from the
combined periods with a local background, which contained a useful time of
$14\ks$ for the PN detector, and $18\ks$ for each of the two MOS
detectors. 

The \Chandra observation was performed with the ACIS-S array exposed, with
the target on the S3 chip. Only standard data preparation was required, as
there were no significant background flares during the short exposure. 

\section{Imaging analysis}
\begin{figure}
\begin{center}
\scalebox{0.4}{\includegraphics*{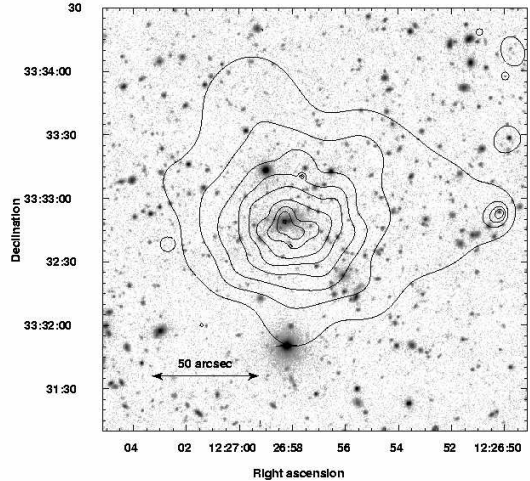}} \\
\caption{\label{fig:xmmoverlay}Contours of X-ray emission detected by \XMM
  $0.3-8\keV$ overlaid on a Subaru I-Band image of cluster
  ClJ1226.9$+$3332. The contours were taken of data from the three cameras
  combined, that was adaptively smoothed so that all features were
  significant at the $99\%$ level. The contours are logarithmically spaced
  above the lowest contour at $0.45\ctpp$.} 
\end{center}
\end{figure}

A combined, exposure-corrected image of the datasets of the PN and two MOS
cameras in the energy band $0.3-8\keV$ was produced, and adaptively
smoothed. Contours of this smoothed emission are shown in
Fig. \ref{fig:xmmoverlay}, overlaid on an optical image. The outer contours
are reasonably circular, suggesting the X-ray emitting gas is fairly
relaxed. The lowest contour, at a level of $0.45\ctpp$ (which is $1.5$
times the background level) is distorted due to the point source in the
West, and truncated slightly along the South-East edge due to a PN CCD gap
that was not fully removed by the exposure correction.

\begin{figure}
\begin{center}
\scalebox{0.4}{\includegraphics*{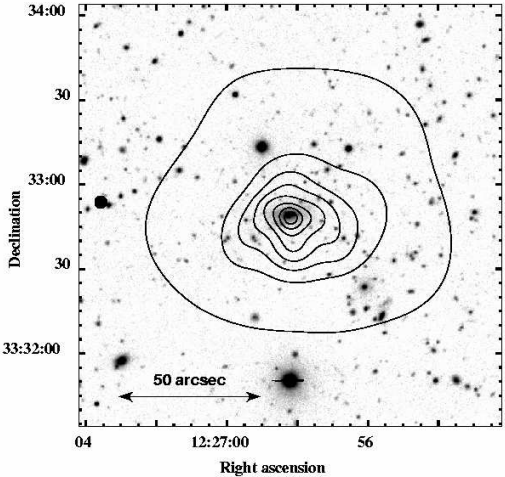}} \\
\caption{\label{fig:choverlay}Contours of X-ray emission detected by
  \Chandra ($0.5-5\keV$) overlaid on the same Subaru I-Band image as Fig
  \ref{fig:xmmoverlay}. The contours are taken from an exposure-corrected
  image that was adaptively smoothed so that all features were significant
  at the $99\%$ level. The contours are logarithmically spaced above the
  lowest contour at $0.03\ctpp$.} 
\end{center}
\end{figure}

For comparison, we also overlay contours produced in the same way from the
archived \Chandra observation of ClJ1226.9$+$3332 on the same optical image
in Fig. \ref{fig:choverlay}. It is clear from Fig. \ref{fig:choverlay} that
there are no strong point sources unresolved in the \XMM observation. 

\subsection{Two-dimensional modelling of the X-ray emission}
A two-dimensional (2D) model of the X-ray emission was fit to the \XMM data, 
taking the different background components and instrumental effects into 
account. The approach followed was to bin the data into
an image with $4.4\arcs$ pixels, but apply no vignetting correction, or any
further manipulation of the data. This pixel size was chosen so as to be an
integer multiple of the $1.1\arcs$ pixels of the point spread function (PSF) 
images produced by the
SAS tool \emph{calview}, allowing them to be re-binned to the same scale,
and to be large enough to reduce computing time in the fitting procedure,
without losing resolution. An image of a dataset obtained with the filter in the closed position (and thus blocking all X-rays) was filtered
in the same way as the source data, and normalised to it, using the ratio
of outside field-of-view counts in the two sets, creating a `particle
image'. This was smoothed with a Gaussian of $\sigma=20\arcs$ to prevent
the fitting being biased by noise, while maintaining any larger scale
spatial variation of this background component. The particle image was then
divided by an exposure map, giving an anti-vignetted image of the particle
component of the background. The exposure map was also used to make a
binary filter mask to exclude the CCD gaps and bad pixels from the fit.  

In a background region of the data (on the same CCD where possible), a
model comprised of the anti-vignetted particle image in that region, plus a
flat component to represent the X-ray background, were multiplied by the
exposure map, convolved with the PSF, and fit to the data, with both
background component normalisations free to vary. This meant that in
effect, the background was fit with a {\it flat} particle background, and a
{\it vignetted} X-ray plus soft-proton background. We note that the
best-fitting normalisation of the particle background varied by less than
$5\%$ from its initial value. This indicates that the normalisation to the
outside field-of-view events was accurate, and therefore the
vignetted-background level found here should also be accurate. These
background normalisations were then fixed, and the source was modelled with
the anti-vignetted particle image in the source region, plus a flat X-ray
background, plus a 2D $\beta$-profile\footnote{http://cxc.harvard.edu/ciao2.3/download/doc/\\sherpa\_html\_manual/refmodels.html}, all multiplied by the exposure map
and convolved with the PSF.  

This procedure was followed for the PN and MOS cameras, and then the fits
were performed simultaneously, with each of the three models using their
appropriate exposure map, fitted background levels, and PSF (images of the
PSF of each telescope were generated at $1.5\keV$, corresponding to the
peak effective area, and at an appropriate off-axis angle). The amplitudes
of the models were independent, but they were constrained to fit to the
same slope, core radius, central position, ellipticity, and rotation
angle. The best-fitting model had a core radius
$r_c=14.5^{+1.2}_{-0.8}{\arcs}$, a slope $\beta=0.66^{+0.02}_{-0.02}$, and
an ellipticity of $0.14$ (while all parameters were free to vary in the
error computation, errors were only computed on $r_c$ and $\beta$ because
of the computational load involved). The fitting was repeated with the PN
and combined MOS data separately, and the best-fitting parameters were
found to be consistent throughout. 

\subsection{One-dimensional surface-brightness profile}
In order to measure the extent of the emission, and to investigate the goodness-of-fit 
of the 2D model to the data, a one-dimensional (1D) surface-brightness profile 
of the emission in an exposure corrected, combined image from the three \XMM 
EPIC cameras was produced. Before the exposure correction, the exposure maps
were normalised to their value at the cluster centroid, thereby
maintaining, as much as possible, the Poissonian statistics in an
exposure-corrected image. 

The profile was centred on the X-ray centroid ($\alpha[2000.0]=12^{\rm
  h}26^{\rm m}57.94^{\rm s}$, $\delta[2000.0]=+33^{\circ}32\arcm 46.2\arcs
$), and the circular radial bins were adaptively sized so that each
contained a detection with a signal-to-noise ratio of at least 3 (the
background level being estimated from a large concentric annulus - we note
that this is likely to be an overestimate of the background at the cluster
centre, due to the anti-vignetted particle background). The emission was
detected out to $100\arcs$ ($776\kpc$) at the $3\sigma$ level.

\begin{figure}
\begin{center}
\scalebox{0.4}{\includegraphics*{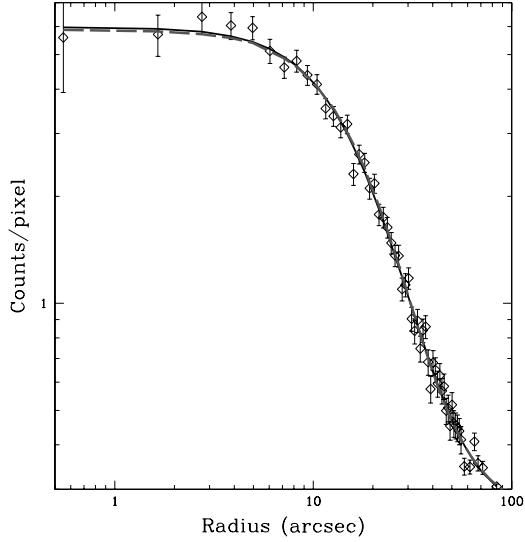}} \\
\caption{\label{fig:radial}Adaptively binned \XMM surface-brightness
  profile of ClJ1226.9$+$3332, with each bin containing a signal-to-noise
  ratio of at least three. The lines show the best-fitting 1D model to the data
  (solid line) and two-dimensional model (dashed line).} 
\end{center}
\end{figure}

The 2D analysis is superior to the 1D analysis, not least because we do not account for the PSF in the 1D analysis. We can however test the goodness of fit of the 2D model in the following way. A 1D profile of the best-fitting 2D model convolved with the PSF was made, and compared 
to the observed 1D profile. A 1D $\beta$-model \citep{cav76} plus a flat background
was fit to both the profile of the data and the 2D model and the best-fitting 
parameters were in good agreement. In the fit to the data, the reduced $\chisq$ 
was $1.09$ for $53$ degrees of freedom. The best-fitting 1D models to the data 
and 2D model are overlaid on a profile of the data in fig \ref{fig:radial}. These 
comparisons indicate that the 2D model provides a good description of the data.

\subsection{\Chandra analysis and the central region}
The archived \Chandra observation of ClJ1226.9$+$3332 was also subjected to
a similar 1D and 2D analysis, and the best-fitting model parameters were
consistent with those derived from the \XMM data, but of lower statistical
precision. 

Many relaxed clusters show excess emission above a $\beta$-model due to cool, 
dense gas in central regions, previously referred to as a cooling flow \citep{fab94b}.
The residuals of the \XMM and \Chandra data after subtraction of the
best-fitting 2D models were examined, and while both showed a weak
central excess, these features were not statistically significant. Excluding the central regions ($r<5\arcs$ for \emph{Chandra}, $r<20\arcs$ for \emph{XMM-Newton}, consistent with the PSF) in the profile fits also gave no significant change to the best-fitting model parameters.

\subsection{Hardness-ratio Mapping}
The temperature structure of the cluster was probed with hardness-ratio
mapping. The hardness ratio, HR, was defined as 
\begin{eqnarray}
HR & = & \frac{H-AH_{bg}}{S-AS_{bg}},
\end{eqnarray}
where $H$ and $S$ are the counts in the source region in the hard and soft
band respectively, and the $bg$ subscript indicates the counts found in a
background region. $A$ is the ratio of the area of the source region to the
background region. Assuming that the errors on each pixel are Poissonian and uncorrelated, 
the error on the hardness ratio is then given by  
\begin{eqnarray}
\sigma(HR)^2 & = & \frac{H+A^2H_{bg}}{(S-AS_{bg})^2} + \frac{(H-AH_{bg})^2(S+A^2S_{bg})}{(S-AS_{bg})^4}.
\end{eqnarray}
A soft band of $0.3-1.1\keV$, and a hard band of $1.1-8\keV$ were chosen
when computing the ratios because these band had similar numbers of net
counts. Images of the cluster emission produced in these hard and soft
bands were binned up adaptively, in order to maximise the signal to noise
in each bin while maintaining good resolution. The minimum number of
background-subtracted counts ($0.3-8\keV$) was set to $150$ per bin,
although a few bins were allowed to fall below this threshold to improve
the resolution. The resultant images were then divided to give a
hardness-ratio map. A series of absorbed \MEKAL spectra were simulated at
different temperatures (assuming a constant Galactic absorption of
$1.38\times10^{20}\pcmsq$ \citep{dic90}, and fixed metallicity of
$0.3\Zsol$), convolved with the appropriate instrument responses, and the
number of counts in the hard and soft bands were found. This enabled the
conversion between HR values and approximate temperatures.  

\begin{figure}
\begin{center}
\scalebox{0.4}{\includegraphics*{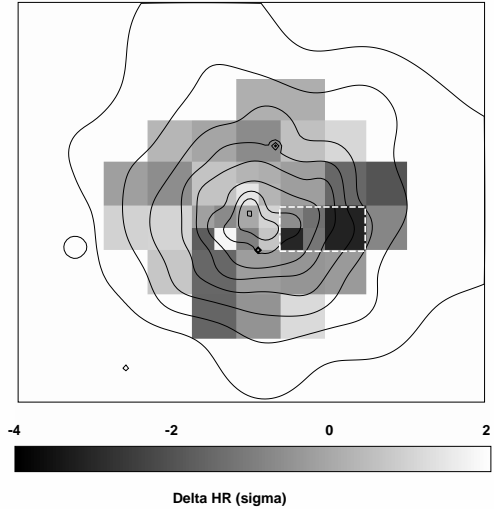}} \\
\caption{\label{fig:HStsig}Adaptively binned HR significance map of the
  \XMM data. The overlaid contours are the same as in
  Fig. \ref{fig:xmmoverlay}. The dashed box contains regions of
  significantly lower temperature than the mean, assuming no variation in
  absorbing column.} 
\end{center}
\end{figure}

Fig. \ref{fig:HStsig} is an image of the differences between the HR in each
bin from the HR corresponding to the global spectrally-measured temperature
($11.5\keV$; see \textsection \ref{sect:specres}) divided by the errors on
both the local HR and the HR of the global temperature added in
quadrature. Pixels where the broadband net counts were $<50$ are excluded,
and the remaining pixels have an average of $140$ counts. This significance
map shows that, within the limits of the data, the emission is generally
isothermal; $66\%$ of the pixels are within $1\sigma$ of the HR
corresponding to the global temperature, and $95\%$ are $<2\sigma$ from
this HR. 

A region of significantly cooler emission to the west of the cluster centre
is marked with a white, dashed box in Fig. \ref{fig:HStsig}. The two
darkest pixels here are just over $3\sigma$ softer than the global
temperature HR (note that we would expect only $0.12$ pixels to be
$>3\sigma$ from the mean if they were randomly distributed). Spectra were
extracted within this region, and fit with an absorbed \MEKAL model. The
best-fitting temperature was $6.5^{+1.2}_{-1.0}\keV$ with the abundance
frozen at $0.3\Zsol$, or $5.9^{+1.1}_{-0.9}\keV$ with a poorly constrained
abundance of $0.8\Zsol$. The reduced $\chisq$ in both cases was $1.3$ for
$29$ (or $28$) degrees of freedom, suggesting that, though the statistical
errors are large, a simple \MEKAL model is not a good description of the
emission from this region (it is excluded at the $88\%$ level). The
\Chandra data indicate that point source contamination is unlikely; a more
likely cause is multi-temperature gas in this region. A plausible
explanation is that we are observing the in-fall of some cooler ($<6\keV$)
body, whose emission is mixed with that from the hotter gas along the line
of sight. 
 
\section{Spectral analysis methods}
When performing spectral analysis, one must be particularly careful to
treat the background components correctly, as failing to do so can strongly
influence the results. We have performed a thorough investigation of
different methods of treating the background spectra, which are in general
extracted in one of two ways. One may extract a local background spectrum
from a large region of the same CCD as the source emission. This method has
the disadvantage that instrumental features in the background spectrum and
the response of the detectors vary across the CCD, and the background
spectrum will be more severely vignetted than the source spectrum, tending
to come from further off-axis.  

Alternatively, one may use a background spectrum extracted from a blank-sky
dataset, which is a combination of several observations with all bright
sources removed. This method has the advantage that the spectrum can be
extracted from the same detector region as the source spectrum, and that
the effective exposure time of the blank-sky dataset can be many times
longer than that of the observation, reducing the Poissonian errors on the
background spectrum. The disadvantages of this method are that the
blank-sky observations are taken at different times and pointings to the
source data, and the background varies both directionally and
temporally. In particular the soft X-ray background varies directionally
due to absorption in the galaxy, and emission from the local bubble, while
the shape and amplitude of the non-X-ray background spectrum varies
temporally due to soft-proton flaring events, and variations in the
particle flux. 

In principle, one can compensate for the vignetting of the telescope by
weighting each event (using SAS 5.3's \emph{evigweight}). The weight is
derived from the ratio of the effective area at the position and energy of
each event, to the effective area at that energy on-axis. This method is
described in detail by, for example, \citet{arn02b}. The disadvantage of
applying this weighting is that non-vignetted particle induced events are
also weighted, artificially boosting their contribution. This effect can be
avoided when using a blank-sky background because the source and background
spectra are extracted from the same detector region. Providing that the
particle contribution is the same in the source and blank-sky datasets,
then the particle weighting effect will be the same, and its effect will
cancel when the spectra are subtracted. The spectra produced from these
weighted datasets should resemble the spectra one would detect with a flat
detector, so the on-axis Ancillary Response File is used when performing
the spectral fitting. 

Thus four spectral background methods were investigated; local and
 blank-sky backgrounds, with and without weighting. These methods were
 applied to the low background and high background periods (see
 \textsection \ref{sect:dataprep}), and both periods combined. In each
 case, the source spectrum was extracted from a circle of radius $100\arcs$
 centred on the cluster centroid. We note that this region crosses a PN CCD
 gap, but the responses of the PN CCDs are identical, and do not vary
 strongly across the chip, so this should not present a significant source
 of uncertainty. The spectra were all fit with an absorbed \MEKAL model, in
 the range $0.3-8\keV$, with abundances fixed at $0.3\Zsol$ and the
 absorbing column density fixed at the Galactic value of
 $1.38\times10^{20}\pcmsq$ \citep{dic90}.

\subsection{The Low Background Period}
All of the background methods described below gave temperatures consistent
with $11.5\pm2\keV$. We take this as a reliable measurement of the
temperature, free of systematic errors, but now check the results when, in
addition, the high background period is included. 

\subsection{Local background, no weighting}
This method is the most straightforward, and given the high background
level in this dataset, is likely to be the most reliable. A background
spectrum was extracted from a large region of the same CCD, at
$\approx250\arcs$ from the cluster centre. This was far enough to avoid
contaminating emission, but as close as possible to reduce the difference
in effective area between the source and background regions. The
best-fitting temperature was $T=11.56\pm1.26\keV$ with a reduced
$\chisq/dof=0.93/298$. We also investigated the dependence of the result on
the background region chosen, by using two other background regions, and
the best-fitting temperatures were all consistent within their $1\sigma$
errors. 

\subsection{Local background, with weighting}
This method is similar to the preceding one, except the spectrum is
produced from weighted events, as described above. This method should
reduce the discrepancy between the effective area at the source region and
the background region. However the contribution of particle induced events
will be incorrectly boosted, and will be boosted more strongly in the
background region which is further off-axis.  

We extracted weighted source and background spectra from the same regions
used above. The best-fitting temperature was $T=4.44\pm0.49\keV$ (reduced
$\chisq/dof=1.15/340$), significantly lower than that found with the
non-weighted spectrum above. This suggests that the anti-vignetting of the
particle events, combined with their high level, has a strong effect in
these data. 

\subsection{blank-sky background, no weighting}
This method uses a spectral background taken from the same detector region
as the source spectrum, in a blank-sky dataset. The high background level
in the source dataset means that its spectrum may be quite different from
that of the blank-sky dataset. We attempt to account for this with the
following method, based closely on that used by \citet{arn02b}. Briefly, 
the blank-sky background was scaled to the data using the ratio of the count 
rates in the whole field of view in the $12-14\keV$ band. A spectrum 
obtained from a background region of the data was subtracted from a corresponding
blank-sky spectrum to produce a `residual spectrum'. This was subtracted from the 
blank-sky spectra to account for systematic residuals between the data and
the blank-sky spectra. Generally, the residual spectrum is taken from further 
off-axis than the source, so will be more
strongly vignetted. This means that any soft X-ray excess (or decrement) in
the source data will be underestimated (or overestimated) to some extent.

Spectra produced with this method were fit as before, giving a temperature
of $T=11.48\pm1.45\keV$ (reduced $\chisq/dof=0.96/298$), in excellent
agreement with the temperature found with the non-weighted local background
method above ($T=11.56\pm1.26\keV$). 

\subsection{blank-sky background, with weighting}
The problem of the vignetting of the residual spectrum in the preceding
method can be solved, in theory, by applying the weightings defined above
to the source and blank-sky datasets, before following the method described
in the preceding section. Again, the particle induced events will be
artificially boosted by the weighting, but in this case, as the source and
background spectra are extracted from the same regions, the boosting factor
should be the same, and it will cancel, providing that the particle event
level in the source and blank-sky sets are similar. Weighted spectra were
produced, following the method above, and the best-fitting model had a
temperature of $T=7.79\pm1.12\keV$ (reduced $\chisq/dof=1.15/340$). This is
not consistent with the temperature found by the two non-weighted methods,
suggesting again that the boosting of the particle events is a significant
effect.  

\subsection{Spectral analysis - summary and conclusions}\label{sec:specsum}
When applied to the low background data, all spectral analysis methods gave
a temperature consistent with $11.5\keV$, with $1\sigma$ errors of
$\approx\pm2\keV$. We believe that this consistency between the methods is
due to the lower particle background in this period. In both the high
background period, and combined periods, the results were consistent with
the low background period when no weightings were used. We believe that the
inconsistencies that emerged when weighting methods were used was due to
the boosting of the higher particle levels in these data. In a further
test, the absorbing column was allowed to vary, along with the temperature,
in our analysis of the combined period data. The Galactic value at the
position of ClJ1226.9$+$3332 is $1.38\times10^{20}\pcmsq$ \citep{dic90};
the best-fitting value with a local background spectrum was
$1.6\pm0.7\times10^{20}\pcmsq$ ($T=11.33\pm1.55$), while with a blank-sky
background spectrum, the best fit was $5.0\pm1.0\times10^{20}\pcmsq$
($T=9.05\pm1.18$). This again shows the reliability of the local background
method. All further analysis was performed on the combined period data,
with a local background as this approach gives the best compromise between
limiting systematic and statistical sources of uncertainty for these
data. The non-weighted blank-sky method was used as a consistency check. 

\section{Spectral results}\label{sect:specres}
The results of the fits to various combinations of the three \XMM cameras
are given in table \ref{tab:spec_results}. All quoted results were found
using a local background with no weighting, though in each case, consistent
results were found using a blank-sky background. All spectral fits for all
combinations of cameras gave consistent results. The spectra were fit in
the $0.3-8\keV$ band, though we note that consistent results were also found 
when fitting in the $1-7\keV$ band.

\begin{table*}
\begin{center}
\begin{tabular}{|c|c|c|c|} \hline 
Camera & $T (keV)$ & Reduced $\chisq/dof$ \\ \hline
PN & $11.56\pm1.26$ & $0.93/298$ \\
MOS1 & $12.42\pm2.80$ & $0.98/98$ \\
MOS2 & $11.93\pm1.76$ & $0.98/103$ \\
MOS1+MOS2 & $12.28\pm1.89$ & $0.99/203$ \\
PN+MOS1+MOS2 & $11.55\pm0.86$ & $1.00/503$ \\
\end{tabular}  
\small{\caption{\label{tab:spec_results}Summary of the results of spectral
    fits to different combinations of the \XMM cameras with a local
    background and no weighting. The temperatures quoted were derived from
    spectral fits with abundances frozen at $0.3$ solar, and the absorbing
    column frozen at the Galactic value.}} 
\end{center}
\end{table*}

\begin{figure}
\begin{center}
\scalebox{0.33}{\includegraphics*[angle=270]{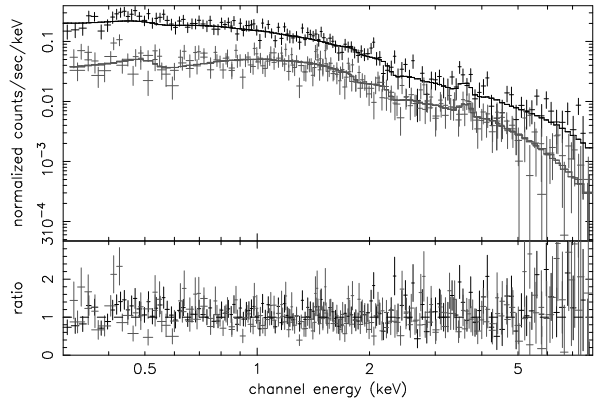}} \\
\caption{\label{fig:spec}PN (upper) and MOS (lower) spectra with the
  best-fitting model. The ratio of data to model values is shown in the lower
  panel. A local background spectrum was used.}
\end{center}
\end{figure}

The simultaneous fit to the data from all three cameras was then
investigated in more detail, with the abundance as a free parameter. The
best-fitting model was $T=11.5^{+1.1}_{-0.9}\keV$ and
$Z=0.33^{+0.14}_{-0.10}\Zsol$ (reduced $\chisq=1.07$ for $502$ degrees of
freedom); this abundance is well constrained for a high-redshift cluster,
and is in good agreement with that found in local clusters (the blank-sky
method gave an abundance of
$Z=0.37^{+0.17}_{-0.17}\Zsol$). Fig. \ref{fig:spec} shows the best-fitting
PN and MOS spectra, produced using a local background. The spectra were
grouped so that each bin contained a minimum of 50 counts (PN) or 20 counts
(MOS).  

The flux of ClJ1226.9$+$3332 measured by \ROSAT in the $0.5-2\keV$ passband
was $3.4\pm0.3\times10^{-13}\flux$. For comparison, the \XMM flux in this
band was $3.7^{+0.1}_{-0.1}\times10^{-13}\flux$ (after extrapolation to
$\r200$). 

\subsection{Temperature Profile}\label{sect:tprof}
\begin{figure}
\begin{center}
\scalebox{0.6}{\includegraphics*{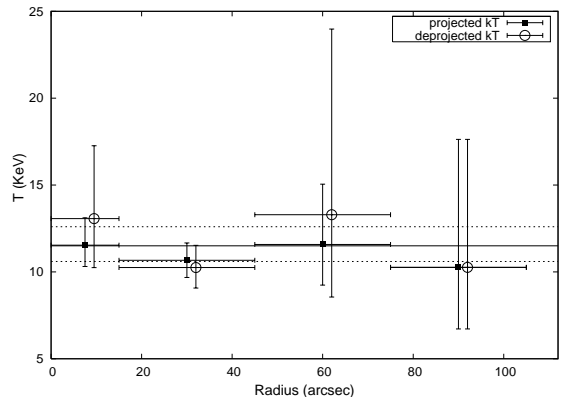}} \\
\caption{\label{fig:tprof}Temperature profile of ClJ1226.9$+$3332, based on
  spectra fit with abundance frozen at $0.3\Zsol$, and a locally extracted
  background. Projected and deprojected temperatures are plotted, with the 
deprojected points offset by $2\arcs$ for clarity. 
The solid line is the best-fitting global temperature, with
  $1\sigma$ errors represented by the dashed lines.} 
\end{center}
\end{figure}

A temperature profile was created by fitting spectra extracted from annular
bins centred on the X-ray centroid. In order to minimise the effect of the
PSF, while maintaining a degree of spatial resolution, the annuli were
chosen so that their width (or diameter in the case of the innermost bin)
were $\ge15\arcs$, which corresponds to the $70\%$ encircled energy radius
of the PSF. Spectra were fit as before in each of these annular bins,
freezing the abundance at $0.3\Zsol$ and the column density at the Galactic
value, using a local background, and fitting in the $0.3-8\keV$ band. The
temperature profile is shown in Fig. \ref{fig:tprof}. The profile is
consistent with isothermality, albeit with large errors, and shows no sign
of any central cool gas.  

The effect of the projection of the emission from the gas in the outer
annuli was then modelled with an `onion skin' method. The temperature
structure was modelled as a series of spherical shells (each of which was
isothermal), and the spectra were fit from the outermost shell in. The
spectrum of a shell was modelled with a single temperature \MEKAL
component, plus a \MEKAL component for each external shell, whose
temperature was fixed at the value measured in that shell, and whose
normalisation was multiplied by a factor. These factors accounted for the
volume of each external shell along the line of sight to the shell being
fit, and the variation in density across each external shell using the
measured gas density profile. This deprojection procedure had no
significant effect on the form of the temperature profile, and did not
reveal any central cool gas, although the size of the errors was increased,
as one would expect, as there were less photons available to constrain the
temperature of the free component in the interior bins. 

\subsection{Entropy Profile}\label{sect:eprof}
The measurement of the gas entropy in groups and clusters of galaxies has
provided evidence for some form of non-gravitational heating
\citep[\egc][]{pon99,llo00,pon03}. In particular, if the entropy profiles
of all systems are scaled by temperature, then cooler systems have a higher
scaled entropy than hotter systems. This contrasts with the predictions of
self-similar models, which include only gravitational heating, where all
scaled-entropy profiles are identical. This indicates that
non-gravitational heating has an impact in cooler systems where it provides
a significant fraction of the gas energy, while its effect is not
detectable in hotter systems. One would expect then, that an extremely hot
system such as ClJ1226.9$+$3332 would have a similar entropy profile to
other hot systems, and our temperature profile of this system allowed a
rare opportunity to measure an entropy profile at high redshift. 

\begin{figure}
\begin{center}
\scalebox{0.6}{\includegraphics*{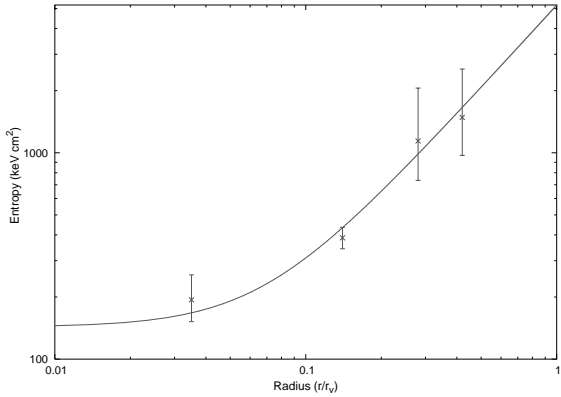}} \\
\caption{\label{fig:eprof}Gas entropy of ClJ1226.9$+$3332 versus radius as
  a fraction of $\r200$. The gas was assumed to be isothermal at $11.5\keV$,
  and the data points give the entropies computed from the measured deprojected
  temperatures shown in Fig. \ref{fig:tprof}.} 
\end{center}
\end{figure}

For consistency with other work \citep{pon99,llo00,pon03}, we defined a
pseudo-entropy, 
\begin{eqnarray}
S & = & T/n_e^{2/3}\ent.
\end{eqnarray}
It was then straightforward to produce the entropy profile shown in
Fig. \ref{fig:eprof}, using the gas density determined from the
surface-brightness profile. The entropy was calculated assuming gas
isothermality at $11.5\keV$, and the data points show the entropies derived
from the measured temperatures in the projected temperature profile. 

It is interesting to note that the entropy observed at $0.1\r200$
($\approx300\pm40\ent$) is significantly lower than that found in local systems
of similar temperature at this radius. For example \citet{pon03} find an
entropy of $\approx550\pm50\ent$ in local systems above $10\keV$. This lower
entropy could be explained by an underestimate of the temperature of
ClJ1226.9$+$3332, however the temperature required to bring the entropy in
line with the local systems is $\approx17\keV$, which seems unlikely. On
the other hand, the central electron density could be overestimated
here. The measured value of $n_e$ was $0.0228\pm0.001\pcmsq$, and a reduction
of $\approx50\%$ is required to bring the entropy at $0.1\r200$ in line
with local values. As discussed in Section \textsection \ref{sect:glob}, the value of $\r200$ used here is subject to systematic uncertainties due to the assumptions made in extrapolating the mass profile. However, these tend to lead to an overestimate of $\r200$, giving an overestimate of the entropy at $0.1\r200$ so this is unlikely to be the cause of the difference between the entropy in local systems and that observed here.

An alternative explanation is that we are observing entropy evolution,
driven by the increasing density of the universe with
redshift. Assuming simple self-similar scaling, the mean density
within a given overdensity radius (relative to the critical density)
is proportional to $H(z)^2$. 
The electron density then scales with redshift as
\begin{eqnarray}
n_e(z) & \propto & H_0^2E(z)^2,
\end{eqnarray}
where
\begin{eqnarray}
E(z) & = &
(1+z)\left(1+z\Omega_M+\frac{\Omega_\Lambda}{(1+z)^2}-\Omega_\Lambda\right)^{1/2}.
\end{eqnarray}
Assuming that the redshift of observation is similar to the redshift
of formation (or at least, the redshift at which the systems last
virialised after a major merger), entropy, when scaled by system
temperature, should therefore evolve as 
$E(z)^{-4/3}$. If the measured 
entropy in ClJ1226.9$+$3332 at 0.1$r_v$ is scaled by this factor (to
give 588$\pm$78 keV cm$^2$), it 
is consistent with, the local \citep{pon03} value. We note that if the dependence of the density contrast $\Delta_c(z)$ on cosmology and redshift (as described by \citet{bry98}) is included in the redshift-scaling of the ClJ1226.9$+$3332 entropy, its value is slightly higher than, but still consistent with, the local \citep{pon03} value.
This suggests that simple, self-similar arguments may explain ICM entropy
evolution. Future papers will examine the evolution of entropy and other
scaling relations using a sample of high-redshift clusters.

\subsection{\Chandra spectral analysis}
A spectrum was also extracted from the archived \Chandra observation of
ClJ1226.9$+$3332, within a $60\arcs$ radius circle, and with a
background extracted from a large concentric annular region of the S3 chip
(excluding point sources). The quantum efficiency (QE) degradation suffered by
\Chandra since launch can cause significant overestimates of cluster
temperatures if not modelled correctly \citep[\egc][]{mau03a}. To account
for this, the \Chandra spectrum was fit with an absorbed \MEKAL model,
including an extra
ACISABS\footnote{http://www.astro.psu.edu/users/chartas/xcontdir/xcont.html}
absorption component. The observation of ClJ1226.9$+$3332 was taken $376$
days after launch. There are also uncertainties in the cross-calibration of the quantum efficiency of the front-illuminated (FI) and back-illuminated (BI) CCDs. It was initially thought that the QE curves were overestimated at low energies by $\approx7\%$ for the FI chips\footnote{http://cxc.harvard.edu/cal/Links/Acis/acis/Cal\_prods/qe/12\_01\_00/}. A more recent reanalysis of pre-flight data has shown that the QE curves of the BI chips are underestimated by $\approx9\%$\footnote{http://cxc.harvard.edu/cal/Links/Acis/acis/Cal\_prods/qe/}. However, due to an additional (as yet unreleased) correction that is required to the telescope effective area, the current best advice for measuring an accurate temperature using the back-illuminated S3 chip is not to apply any additional QE correction. Accordingly, none was applied, but we note that systematic uncertainties at the $\approx10\%$ level may exist. 

The fits were performed in the
$0.6-8\keV$ band, with the column density frozen at the
Galactic value, and the abundance at $0.3\Zsol$. The best-fitting model 
temperature was $12.6^{+3.0}_{-2.2}\keV$, in good agreement with that measured
by \emph{XMM-Newton}. The best-fitting \Chandra temperature was also found
to be consistent when the spectrum was fit in the $1-8\keV$ band, where the 
effects of the quantum efficiency degradation are less severe. 

The unabsorbed flux measured by \Chandra ($0.5-2\keV$) was
$3.6\pm0.1\times10^{-13}\flux$ (after extrapolation to $\r200$), which is
consistent with that measured by \XMM and \ROSAT.  This shows again that
point source contamination was not a problem in the \XMM data. 

\section{Determination of Global Properties}\label{sect:glob}
We have derived the luminosity, gas mass, total mass, and gas mass fraction within two different radii. The most reliable results are those obtained within the extent of the data ($r=100\arcs$, corresponding to an overdensity $\Delta\approx1000$). The easiest results to compare with theoretical models of cluster growth are those extrapolated by a factor of $2.2$ in radius to $\r200$, but systematic uncertainties may be associated with the extrapolation.

The method used was similar (but not identical) to that described in \citet{mau03a}. Briefly, if the gas density profile is described by
a $\beta$-model, then under the assumptions of isothermality, hydrostatic
equilibrium, and spherical symmetry, the total density profile of the
cluster is given by 

\begin{eqnarray}
\bar{\rho}(<r) & = & \frac{M(<r)}{4/3\pi r^{3}}\\ 
& = & 2.70\times10^{13}\beta\frac{T}{\keV}\left(\frac{r}{\Mpc}\right)^{-2}\times\nonumber \\ 
&   & \frac{(r/r_c)^2}{1+(r/r_c)^2}M_{\odot}\Mpc^{-3}.\label{egn:densprof}
\end{eqnarray}

Here, we have adopted a value of $0.59m_p$ for the mean molecular weight of
the gas, where $m_{\rm p}$ is the proton mass. This density profile was
used to estimate $\r200$, and the measured flux was extrapolated out to
this radius, and converted to a luminosity. The central gas density was
computed from the measured \MEKAL normalisation, and the measured gas
density profile was integrated to give the gas mass. The total gravitating
mass within $\r200$ was derived from Eqn. \ref{egn:densprof}. 

The errors quoted on all non-observed quantities were derived from $10,000$
randomisations of the measured quantities under the Gaussians described by
their measured $1\sigma$ errors. The properties of ClJ1226.9$+$3332 are
summarised in Table \ref{tab:summary}. Our assumption of isothermality is
supported by the measured temperature profile, and hardness-ratio mapping,
while the relaxed appearance of the X-ray emission, and the good fit of an
isothermal $\beta-$model to the data indicate that the gas is close to
hydrostatic equilibrium. 

The extrapolation of the cluster properties out to large radii introduces systematic uncertainties which are not taken into account in the above method. In a sample of 66 systems with measured temperature profiles \citet{san03} found that the incorrect assumption of isothermality leads to an average overestimate of $M_{200}$ by $\approx30\%$ and $\r200$ by $\approx20\%$. The overestimation of $\r200$ leads in turn to an overestimation of $M_{gas}$ by $\approx25\%$ at that radius (the $\r200$ and $M_{gas}$ uncertainties were provided by Sanderson (private communication)). These are taken as reasonable indications of the systematic uncertainties on those properties, and are added in quadrature to the statistical errors derived above in the quoted values of these properties.

We find a virial radius of $\r200=1.66\pm0.34\Mpc$ for cluster
ClJ1226.9$+$3332. This means that the properties of the system are directly
measured out to $\approx0.45\r200$. Assuming an extrapolation of the
surface-brightness profile is valid, it is interesting to note that while
this radius encloses $90\%$ of the X-ray emission, it encloses only
$\approx45\%$ of the gas mass and total mass of the system. 

\begin{table*} 
\scalebox{0.8}{
\begin{tabular}{|c|c|c|c|c|c|c|c|c|c|c|} \hline 
$\Delta$ & Redshift & $T (keV)$ & $L_{bol} (erg\s^{-1})$ & $\sigma (\kmps)$ & $r_c ({\kpc})$ & $\beta$ & $r_\Delta (Mpc)$ & $M_{gas} (M_\odot)$ & $M_{tot} (M_\odot)$ & $f_{gas}$ \\ \hline 
1000 & $0.892$ & $11.5^{+1.1}_{-0.9}$ & $4.8\pm0.1\times10^{45}$ & $997^{+285}_{-205}$ & $113^{+9}_{-6}$ & $0.66^{+0.02}_{-0.02}$ & $0.73\pm0.04$ & $6.5\pm0.4\times10^{13}$ & $6.1^{+0.9}_{-0.8}\times10^{14}$ & $0.11\pm0.02$ \\ 
200 & $$ & $$ & $5.3\pm0.2\times10^{45}$ & $$ & $$ & $$ & $1.66\pm0.34$ & $1.7\pm0.4\times10^{14}$ & $1.4\pm0.5\times10^{15}$ & $0.12\pm0.05$ \\ \hline 
\end{tabular}
}
\caption{\label{tab:summary}Summary of the measured and inferred properties
  of cluster ClJ1226.9$+$3332 based on \XMM observations, assuming a
  cosmology of $\Omega_{M}=0.3$ $(\Omega_\Lambda=0.7)$ and
  $H_0=70$\kmpspMpc. The first line gives the properties within the detection radius, corresponding to an overdensity of $\Delta=1000$. The second line gives the properties when extrapolated to an overdensity radius of $\Delta=200$.}

\end{table*} 

\section{Velocity Dispersion}
ClJ1226.9+3332 was observed by us on April 18, 2002 with the LRIS
spectrograph \citep{oke95} on the Keck-I 10m telescope. We used
the 600 l/mm grism blazed at $1\micron$, and a multi-object spectroscopy
mask with 1.25'' wide slits. Further details of the observational
setup and the data reduction procedure will be provided in a future
paper \citep{ebe03}. From 12 accurately
measured cluster redshifts (individual radial velocity error less than
30 km s$^{-1}$) and using a biweight estimator for the systemic
cluster redshift $z$ and the comoving cluster velocity dispersion
$\sigma$ we find $z=0.8920$ and $\sigma=997^{+285}_{-205}\kmps$
using the ROSTAT statistics package \citep{Bee90}.

The observed velocity dispersion is consistent with the 
measured X-ray temperature, given the scatter in the local $T-\sigma$
relation of \citet{xue00b} 
The velocity histogram, although poorly constrained with only
12 velocities, shows no signs of significant substructure.

\section{Discussion}
ClJ1226.9$+$3332 is the highest temperature galaxy cluster known at
$z>0.6$, and, uniquely at these redshifts,
 is an extremely massive system (similar in mass to the Coma cluster)
 which appears to be relaxed. Images of both the \XMM observation
 analysed here, and the archived \Chandra observation show almost
 circular isophotes, and no obvious large-scale substructure. Within
 the limits of the current data, the cluster is generally isothermal
 (except for one small cooler region). The relaxed nature is
further supported by the good agreement of the $\beta$-model with the
surface brightness distribution.  
This relaxed appearance is important in justifying the assumptions
used to derive the total mass. 

The existence of even one high-redshift cluster of this mass can be used to constrain cosmological models. We initially test for consistency with the \LCDM cosmology of \citet{spe03} from Wilkinson Microwave Anisotropy Probe (WMAP) data, using their results based on a model with a constant spectral index of primordial fluctuations. In this cosmology, at a redshift of $0.89$ we expect to see a density of systems more
massive than CLJ$1226.9+3332$ of $4.86\times 10^{-3}$ deg$^{-2}$ per unit
$z$. We have
adopted the \citet{jen01} halo mass function in this calculation,
and converted between our mass definition ($M_{200}$ relative to the
critical density) and that of \citet{jen01}  ($M_{180}$ relative to
the background density) via: $M_{180}/M_{200}=1.14$, assuming an NFW \citep{nav96}
profile with concentration parameter $c=5$. Given that CLJ$1226.9+3332$
was detectable in the WARPS over the full survey area of 73 deg$^{-2}$ and
to a redshift of $z=1.8$ and (very conservatively) assuming no further
evolution in the cluster mass function beyond $z=0.89$ we would expect a
total of $0.64$ such clusters in the entire survey. If the cluster mass within $\r200$ is $\approx30\%$ lower, as estimated from the combination of systematic and statistical errors, then the predicted number of such clusters rises to 2.4. The detection of one
such cluster is therefore consistent with this model.

Interestingly, the predicted  number reduces to  0.23 in the  running
spectral index  WMAP model  in  which the spectrum of primordial density
fluctuations is  a slowly  changing power law  as a function  of scale
and in which   the third  derivative  of the  inflation
potential plays a role \citep{pei03}.
This model  was invoked  \citep{spe03}
primarily  to  investigate the  apparent  effects  of combining  other
experimental CMB  data with that of  WMAP, in which  the small angular
scale  amplitude  of fluctuations  seem  to  be systematically lower than
the overall best-fit  amplitude.
The   existence  of  CLJ1226  therefore  mildly
disfavours the running index model. However, if the cluster mass is
lower, but still within the measurement errors, then the predicted
number of such clusters rises to 0.86, consistent with observation.
The power of massive  clusters at high  redshift to discriminate
between  cosmologies is illustrated by this example, but a key
requirement is accurate mass measurements from data extending
to the virial radius.

Although no longer a viable model we note for
completeness that the probability of observing a cluster of at least this
mass in a high density ($\Omega_M=1$, $H_0=50$) Universe is approximately
$8\times 10^{-5}$, or $\sim 1/13,000$ (or $\approx2\times10^{-4}$ for a cluster mass at the low end of the measurement errors).

In relaxed clusters, where the central gas cooling time is
sufficiently low, gas may cool to a temperature of $\sim1/3$ of that
of the surrounding gas. The cooling time of the intra-cluster gas was
estimated by dividing its thermal energy by its luminosity in a series
of concentric spherical shells. The \Chandra density profile was used
for this because of its superior resolution, though the results from
\XMM were consistent. The radius within which the cooling time is less
than the age of the universe at the cluster's redshift ($6.22\Gyr$) is
$92\kpc$ ($12\arcs$) in our \LCDM cosmology. There is no significant
central excess emission seen, and the interior bin of the temperature
profile shows no evidence for any cooler gas. The weak residual counts
from the 2D surface brightness fitting were used to estimate that any
central cool gas contributes less than $5\%$ of the cluster luminosity
(assuming a $5\keV$ \MEKAL spectrum for the cool gas). Numerical
simulations have shown that merger events can disrupt central cooling
in clusters \citep[\egc][]{rit02}. A plausible explanation, then, for
any lack of central cool gas is that the system is being observed
after some recent minor merger. While the gas appears to have relaxed
into hydrostatic equilibrium on large scales, traces may remain in the
cooler gas observed to the west of centre, which may be an in-falling 
poor cluster or group. 

The gas mass fraction of ClJ1226.9$+$3332 measured within the spectral
extraction radius of $100\arcs$ was $0.11\pm0.02$, and
$0.12\pm0.05$ when the mass profiles are extrapolated out to the
virial radius. These values are consistent with those seen in local and
intermediate-redshift clusters
\citep{vik99,sad01,all02a,ett03}. \citet{all02a} and \citet{ett03} also use
the apparent variation in $f_{gas}$ with redshift to constrain cosmological
parameters. The measurement of $f_{gas}$ presented here, along with others
at similar redshifts will allow this method to be extended in redshift. 

The metal abundance of $Z=0.33^{+0.14}_{-0.10}\Zsol$ measured in
ClJ1226.9$+$3332 is well constrained for such a high-redshift cluster, and
is typical of values found in local clusters. This measurement is
consistent with the lack of evolution in Fe abundance and high redshift of
enrichment ($z>1$) of the ICM proposed by \citet{mus97a} and recently confirmed by \citet{toz03}. 

Luminous clusters like ClJ1226.9$+$3332, with measured luminosities and
temperatures provide useful tools for calibrating the
luminosity-temperature (L-T) relation at high redshifts. The luminosities
predicted by two local L-T relations for a cluster with the temperature of
ClJ1226.9$+$3332 were compared with the measured luminosity. With the L-T
relation expressed as $L=A(T/6\keV)^B$, \citet{arn99} (hereafter AE99) find
$A=2.88\pm0.20\times10^{44}h_{100}^{-2}\ergps$ ($h_{100}=H_0/100\kmpspMpc$)
and $B=2.88\pm0.15$, which predicts
$L=3.8^{+2.1}_{-1.2}\times10^{45}\ergps$. The L-T relation of
\citet{mar98a} (hereafter M98)
($A=3.11\pm0.27\times10^{45}h_{100}^{-2}\ergps$, $B=2.64\pm0.27$) predicts
a luminosity of $L=3.5^{+2.4}_{-1.3}\times10^{45}\ergps$. The measured
luminosity of ClJ1226.9$+$3332 ($5.3\pm0.2\times10^{45}\ergps$) is higher
than the predicted values, but not significantly so. The L-T relations
above were derived for clusters with weak or absent cooling flows (AE99),
or with cooling flow emission excluded (M98), so it should be reasonable to
compare them with this cluster. The normalisation of the L-T relation (measured within a fixed overdensity radius) is predicted to evolve with redshift, by a factor $E(z)$. The predicted
luminosities, scaled by $E(z)$ in our
\LCDM cosmology ($1.65$), increase to
$6.3^{+3.5}_{-2.0}\times10^{45}\ergps$ (AE99), and
$5.8^{+4.0}_{-2.1}\times10^{45}\ergps$ (M98). These values agree well
with the observed luminosity, although as stated above, the measured
luminosity is also consistent with no evolution. Including the redshift-dependence of the density contrast $\Delta_c(z)$ in the predicted evolution does not affect this result.

The same comparisons were made adopting a cosmology of $H_0=50\kmpspMpc$
and $\Omega_M=1$ $(\Omega_\Lambda=0)$. In this cosmology, the observed
luminosity of ClJ1226.9$+$3332 was $6.1\pm0.2\times10^{45}\ergps$, and
$C(z)=2.594$. The predicted luminosities of both L-T relations agree well
with the observed value without applying the evolution factor. When
evolution is included, the predicted luminosities are
$19.5^{+10.9}_{-6.2}\times10^{45}\ergps$ (AE99), and
$18.7^{+13.0}_{-7.0}\times10^{45}\ergps$ (M98). Thus, in this cosmology the
measured luminosity of ClJ1226.9$+$3332 is inconsistent with the predicted
evolution of the L-T relation, at the $\approx2\sigma$ level. 

\section{Conclusions}
ClJ1226.9$+$3332 is a remarkable and unique cluster.
We have performed a detailed analysis of an \XMM observation, and
after careful comparison of background subtraction methods, we have
confirmed its high temperature, and produced a temperature profile for
the first time at this high redshift ($z=0.89$). The total mass is
found to be extremely high ($1.4\pm0.5\times10^{15}M_\odot$)
and similar to that of
the Coma cluster. The probability of such a cluster being found in the
discovery survey is $0.64$ (assuming a \LCDM cosmology).

The relaxed, and generally isothermal, X-ray appearance, together with 
the gas mass fraction, metal abundance, and gas density profile slope
($\beta$) all being consistent with those of local clusters, suggests 
that this cluster was assembled significantly earlier than z=0.9.

The high luminosity and relaxed nature make it an extremely useful
subject for further studies of the gas, dark matter and galaxy
properties out to large radii at high redshift. Deeper \Chandra and \XMM observations
are planned, in part to test the assumptions of isothermality and
hydrostatic equilibrium which underpin the derivations of many of the
cluster properties. 

\section{Acknowledgements}
We thank Eric Perlman, Pasquale Mazzotta, and Monique Arnaud for
discussions of this work, and Elizabeth Barrett for her work on the
Keck spectroscopy. We thank Zoltan Haiman for his help with cosmological
modelling. The referee made useful comments which improved this paper. BJM is supported by a PPARC postgraduate
studentship. HE and CS gratefully acknowledge financial support from NASA
grant NAG $5-10085$. 

\bibliographystyle{mn2e}
\bibliography{clusters}

\end{document}